\documentstyle[12pt]{article}
\begin{document}
\centerline{\Large \bf Condensation of Hard Spheres Under Gravity:}
\centerline{\Large \bf Exact Results in One Dimension}
\vskip 1.0 true cm
\centerline{\bf Daniel C. Hong}
\vskip 1.0 true cm
\centerline{\bf Department of Physics, Lewis Laboratory}
\centerline{\bf Lehigh University, Bethlehem, PA 18015 \\}
\begin{abstract}
We present exact results for the density profile of the one dimensional 
array of $N$ hard spheres of diameter $D$ and mass $m$ under 
gravity $g$.  For a strictly one dimensional system, the
liquid-solid transition occurs
at zero temperature, because the close- packed density, $\phi_c$,
is one.  However, if
we relax this condition slightly such that $\phi_c=1-\delta$,
we find a
series of critical temperatures, $T_c^{(i)}
=mgD(N+1-i)/\mu_o$ with $\mu_o=1/\delta -1$,
at which the $i$-th particle undergoes
the liquid-solid transition.  The functional form of the onset temperature,
$T_c^{(1)}=mgDN/\mu_o $, is consistent with
the previous result [Physica A 271, 192 (1999)]
obtained by the Enskog equation.  We also show that
the increase in the
center of mass is linear in $T$ before the transition, but it becomes
quadratic in
$T$ after the transition because of the formation of solid near the bottom.
\end{abstract}
\vskip 1.0 true cm
\noindent {\bf I. Introduction}
\vskip 0.2 true cm
In the previous paper [1], the author proposed that the hard sphere gas
undergoes the condensation transition under gravity $g$, 
and identified the transition
temperature, $T_c$, as the point at which the Enskog equation [2] fails to
conserve the total number of particles.  Based on the
fact that hard spheres cannot be compressed beyond the close-packed density,
it was suggested [1] and confirmed [3] that the missing particles should
condense from the bottom and form a solid below $T_c$, and its fraction
in a solid regime at a temperature $T<T_c$ was predicted to be $1-T/T_c$.
The transition temperature $T_c$ was determined as
$$T_c=mgD\mu/\mu_o \eqno (1)$$
where $m, D$ are the mass and the diameter of a hard sphere, $\mu$ is
the layer number of the system, and $\mu_o$ is a constant, which 
depends on the level of approximations in truncating the
BBGKY heirarchy [4], or perhaps in employing the density functional theory [5].
Hence, the
value obtained by the Enskog theory [1], or more precisely 
by the Enskog pressure,
may be close to the real value, 
but not precise.  For example, if one uses the pressure form suggested by 
Percus-Yevick [6], 
this constant will be slightly different.  For a one dimensional
lattice gas [7,8], it can be shown
that $\mu_o=-ln(\alpha/(1-\alpha))$ with $\alpha=exp(-14)$.
The crucial point, however, is that
the scaling form of the transition temperature (Eq.(1)) should survive in
all approximations.  The purpose of this paper is to demonstrate this point
by exactly solving the one dimensional hard sphere problem under gravity.
For a strictly one dimensional system, the condensation transition
occurs at zero temperature, because the close-packed density, $\phi_c$, is one.
However, some useful information may be extracted from 1d results if we 
relax this condition slightly such that $\phi_c=1-\delta$.  Then,
the transtion occurs at a finite temperature.  
Even though the fluctuations in 3d are very different from 1d system,
we will show in this paper that results obtained in this way appear
to be relevant to the real physical system.  Perhaps,
one may view
such a  1d system as a coarse grained mean field system of the real
three dimensional hard sphere system.  We will
obtain the exact transition
temperature, $T_c$, 
and check its functional form against Eq.(1).  We also determine 
the series of transition temperatures, $T_c^{(i)}$, at which the $i$-th layer
undergoes the condensation transition.  We further
show how a sharp departure
in the center of mass statistics shows up before and after the transition.
Before the transition, the increase in the center of mass is {\it linear}
in $T$, while after the transtion it is {\it quadratic}
in $T$, because of the formation
of solid near the bottom, which is a characteristic of Fermi systems [9,10].  

\vskip 0.5 true cm
\noindent {\bf II. Condensation of 
One Dimensional Hard Sphere Gas Under Gravity}
\vskip 0.3 true cm
Consider a collection of hard spheres of finite radius $R$
(or diameter $D=2R$)
in a one dimensional tube with the top open.  Let the mass of the $i$-th
particle be $m_i$.
We assume that each hard sphere is in thermal equilibrium with
a heat reservoir at a temperature $T$. The system we have in mind is
the one used in usual Molecular Dynamics simulations, where each particle
is kicked periodically by Gaussian noise so that the average kinetic
energy of each particle, $m<v^2>/2=T$.  We ignore the pressure due to the
reservoir.  In such a case, since the kinetics is separated out,
we only consider the configurational integral in computing the partition
function, $Z_N$, of the $N$ particle assembly:

$$ Z_N = \int_R^{\infty}dz_1\int_{z_1+2R}^{\infty}.......\int_{z_{N-1}+2R}^
{\infty}dz_N exp(-\beta' g(m_1z_1+......+m_Nz_N)) \eqno (2)$$
with $\beta'=1/T$. The hard sphere gas without gravity has been
studied and is known as the Tonk gas [11].
The integral in (2) involves the
exponential functions and thus can be carried out
exactly to yield,
$$ Z_N ={\frac{1}{(\beta' g)^n}}\bullet {\frac{e^{-2\beta' gm_NR}}{m_N}}
{\frac{e^{-2\beta' g(m_N+m_{N-1})R}}{(m_N+m_{N-1})}}.....
{\frac{e^{-2\beta' g(m_1+m_2+...+m_N)R}}{(m_1+m_2+.....+m_N)}} \eqno (3)$$
We now compute average quantities.  First, 
the average position of the $i$-th particle $<z_i>$ is given by:
$$<z_i> = -{\frac{1}{\beta' g}}{\frac{\partial ln Z_N}{\partial m_i}}
=(2i-1)R + {\frac{T}{g}}\bar z_i \eqno (4)$$
where
$$\bar z_i = \Sigma_{j=1}^{i}({\frac{1}{\Sigma_{k=j}^N m_k}})\eqno (5) $$
If all the masses are the same, i.e. $m_i=m$, then this reduces to

$$<z_i>/D = (i-1/2) + {\frac{T}{mgD}}\Sigma_{j=1}^{i}{\frac{1}{N+1-j}}
\eqno (6)$$
Note that the first term, $z_i(0)/D = i-1/2$, results from
the close-packing at the ground state $T=0$ and the second term
represents the thermal expansion.  Note also that
$\sum_i^N <z_i>/D = N^2/2 + TN/mgD$.  The dimensionless thermal
expansion defined as, $\bar z_i = (<\Delta z_i>/D)(mgD/T)$ with
$<\Delta z_i> = <z_i(t)>-z_i(0)$,
is independent of the temperature.  For example,
$$\bar z_1=1/N, \qquad \bar z_2=1/N + 1/(N-1), \qquad 
 \bar z_N = 1/N+1/(N-1)+....+1/2+1$$

The dimensionless
mean expansion per particle is precisely given by the thermal energy
injected into the system:

$$ <\bar z(T)>= \frac{1}{N}[\sum_i^N<\Delta z_i>]\frac{mg}{T}=\frac{1}{N}
\sum_i^N<z_i(t)-z_i(0)>{\frac{mg}{T}}=
\sum_{i=1}^{N}\bar z_i(T) = 1 \eqno (7a)$$

The change in the center of mass due to the thermal expansion is {\it linear}
in $T$:

$$<z(T)>=\frac{1}{N}\sum_{j=1}^N(<z_i(T)>-z_i(0))>=T/mg\eqno (7b)$$
\vskip 0.3 true cm
We now compute the density profile, $\rho(z_i)$, as a function of
position, $z_i$.  
Define the dimensionless density, $\phi(z_i)
=\rho(z_i)/\rho_c$ with $\rho_c=1/D$. Then, since 
$\rho(z_i)\Delta z_i = \Delta i$, we find
$\phi(z_i) = {\frac{\Delta i}{\Delta z_i}}/\rho_c$ and its 
discrete version becomes:

$$ \phi(z_i) = 1/[1+{\frac{1}{\beta}}{\frac{1}{N+1-i}}] \eqno (8)$$
where we have used the relation:
$$1/N + 1/(N-1)+... + 1/(N-i+1) \approx -\int_{x=N}^{x=N-i+1}dx/x = 
-ln(1-(i-1)/N)\eqno (9)$$
and we have redefined the dimensionless temperature $\beta =mgD/T$ and 
the dimensionless coordinate, $y_i=<z_i>/D$:
$$y_i = (i-1/2) + {\frac{1}{\beta}}\sum_{j=1}^{i}{\frac{1}{N-j+1}} \eqno (10)$$
Note that $\int\rho(z_i)dz_i= \int\phi(y_i)dy_i= N\phi_c$ with $\phi_c$ the
close-packed density.  The density at the bottom layer, $\phi_o$,
is given by Eq.(8) with $i=1$, i.e., $\phi_o=1/[1+1/\beta N]$.
For a strictly one dimensional system, the closed packed density,
$\phi_c=1$, and thus by setting, $\phi_o=\phi_c=1$, we find 
that the crystallization
occurs at zero temperature in one dimension.

In order to extract some useful information from 
one dimensional results, and make them relevant to higher dimension, we
assume that the close-packed density is slightly below one by a small
amount, $0 < \delta\ne 0 <<1$, i.e., $\phi_c=1-\delta$.
What we have in mind is a coarse grained three dimensional
system, for which
each column may interact weakly.  
In fact, we have found that such a system can be realized in the Molecular
Dynamics simulations 
if the system is initially arranged in a two dimensional
square lattice with a little
space between the columns.  In such a case, particles in each column
do not mix, and the square structure is maintained [18].
Such a model could be
understood as a coarse grained mean field model in the spirit of ref.[7]. 
Certainly, the 
fluctuations in $3d$ are very different from those in 
$1d$, and thus it may be
objectionable to extent the results of $1d$ to $3d$.  Nevertheless, 
the results of the $1d$ obtained this way with regard to
the existence of the condensation
temperature, and perhaps the existence of the discrete jump in the
condensation process may survive in high $d$, as will be shown shortly.

Now, if we let $\phi_c = 1-\delta $, then 
one can easily find the onset of the condensation temperature, $T_c$,
at which the
first layer becomes crystallized:
$$ T_c=mgD\mu/\mu_o\eqno (11)$$
where $\mu=N$ is the initial layer number (or the Fermi energy [10]), 
and the constant $\mu_o$ is given by
$$\mu_o=\frac{1}{\delta}-1 \eqno (12)$$
Note that Eq.(11) has the same funtional form as (1).
One may relate $\delta$ to the critical pressure, $P_cD^2
=\mu_o T_c/D$ at which
the crystalization occurs.  From the force balance equation, we find the
pressure at the bottom:
$$ P(0)D^2 = mg\int_o^{\infty}dz\phi(z)= mgN\phi_c \eqno (13)$$
The factor $D^2$ was introduced to effectively model the three dimensional
system.
By equating $P(0)$ to the critical pressure, $P_c$, we again
find the transition temperature, $T_c=mgD\mu/\mu_o$.
Ref.[7] identifies the critical pressure as:
$P_c=14T_c/D^3$.  Hence, we find: $\mu_o=1/\delta-1= 14.937$, and 
$\delta \approx 0.06275$.  
After the first layer becomes cystallized at $T_c$, the density profile
above the second layer is given by Eq.(8)
with $N$ replaced by $N-1$, and $i=1,..., N-1$.  This
is effectively equivalent to shifting the origin from the first to the
second layer.  The second layer, which has now become the origin,
becomes crystallized at the second critical temperature, 
$T_c^{(2)}$:
$\phi_o(T_c^{(2)})=1-\delta$.  
The process continues, and we can find a series of
critical temperatures, $T_c^{(i)}$, at which the
$i$-th layer in the original labeling becomes crystallized:
$$T_c^{(i)}=\frac{mgD(N+1-i)}{\mu_o}\eqno (14)$$ So, all the particles
are crystalized at $T=T_c^{(N)}=mgD/\mu_o$, which is {\it not} the
absolute zero, because $\delta\ne 0$.  Note also that the
crystallization of each layer proceeds with a discreet temperature
jump, $\Delta T=T_c^{(i+1)} - T_c^{(i)} = mgD/\mu_o$.  Hence, the heat
release, or the latent heat, $Q$, resulting from the formation of one
solid layer is $Q = \Delta T = mgD/\mu_o$.  Biben et al [12]
investigated the density profile of a hard sphere suspension in a
gravitational field using Monte Carlo simulations, and reported that
for $\Delta = mgD/T \le 2.5$, the system is a strongly perturbed
fluid, while at $\Delta \approx 2.75$ the first two layers form a
crystal, and the formation of third and fourth layer crystals occurs in
a {\it discontinuous} manner between $\Delta = 2.5$ and $\Delta
=2.75.$ Setting $mgD/T_c=2.5$, and $ T_c=mgD/\mu_o$, we find
$\mu_o=2.5$ and $\delta \approx 0.2857$,  and the gap $\Delta
T=mgD/\mu_o \approx 0.25$.  Such findings do not seem to be
inconsistent with the results presented above.
\vskip 0.3 true cm
We now examine the center of mass statistics below the condensation point 
$T_c=T_c^{(1)}$.  At a given temperature $ T_c^{(i+1)}< T < T_c^{(i)}$,
what is the fraction of particles in a condensed regime?  At this temperature,
particles up to the $i$-th layer are condensed. 
Then, the fraction of particles in the condensed regime, $\zeta_F\equiv i/N$,
which is termed the Fermi surface [1],
is given by a simple manipulation of identities:
$$ \zeta_F = i/N = 1-[\frac{N-i}{N}] = 1-T/T_c\eqno (15)$$
where we used, $ T_c^{(i+1)}/T_c = [mgD(N-i)/\mu_o]/mgDN/\mu_o]= (N-i)/N
\equiv T/T_c$.
\vskip 0.3 true cm
Now, the dimensionless center of mass, $ <y(T)>\equiv <z(T)>/D$, is given by:
$$ <y(T)> = \int_o^{\infty}dy y \phi(y)/\int_o^{\infty}dy\phi(y) 
\equiv I_2/I_1 \eqno (16)$$
where the integral now splits into two due to the formation of a solid below
$\zeta_F$.  More precisely, 
$$I_1=\int_o^{\zeta_F} \phi_c dy + \int_{\zeta_F}^{\infty} \phi(y-\zeta_F)dy
=\phi_c\zeta_F + (N-\zeta_F)\phi_c=N\phi_c \eqno (17)$$
We need some manipulation 
in computing the denominator $I_2$.  To this end, we again
split $I_2$ into two
integrals: one for the solid regime, which is essentially a rectangle, and
the other for the fluid regime, where the density profile is given by (8) but
with N replaced by $N'=N-\zeta_F$.  Hence,
$$I_2=\int_o^{\zeta_F} y\phi_c dy + \int_{\zeta_F}^{\infty} 
y\phi(y-\zeta_F)dy=\phi_c\zeta_F^2/2 + \zeta_F\phi_c(N-\zeta_F) + J\eqno (18)$$
where 
$$J\equiv \int_o^{\infty} dy y\phi(y) = \sum_{j=1}^{N-\zeta_F}y_j\phi_j
(\frac{\Delta y_i}{\Delta i})_{i=j} =  \sum_{j=1}^{N-\zeta_F} z_j/D
\eqno (19)$$
But $\sum_j^{N'}z_j=D[N'^2/2+ TN'/mgD]$(Eq.(6)).  Hence, with $N'=
N-\zeta_F=NT/T_c$, we find:
$$J=[N^2T^2/2T_c^2 + (N^2T^2/\mu_oT_c^2)]=N^2\Lambda(T/T_c)^2 \eqno (20)$$
where $\Lambda = [1/2 + 1/\mu_o]$.
Note that the increase in the center of mass is {\it quadratic} in $T$, namely:
$$<\Delta z(T)> = <z(T)>-ND/2 = \alpha ND (T/T_c)^2 \eqno (21)$$
with $\alpha=[(2+\delta(1-\delta)]/[2(1-\delta)^2]$, which is 
a characeristic of Fermi systems [10]. 
\vskip 0.5 true cm
In passing, we make the following remarks. In ref. [7], an
attempt was made to derive
the condensation point for the lattice gas, which is again consistent with
the form given by Eq.(1).  While the
lattice gas may capture some of the essence of the hard sphere systems,
it is important to recognize that the logarithmic singularity in the
pressure of the lattice gas 
[7,13] is far different from the power law singularity
of the real hard sphere gas [14].  Finally, the relevance of the present
study to granular materials [15]: Granular materials are macroscopic particles,
and the parameter $\Delta= mgD/T \approx 10^{13}$
is an astronomical number, if one uses a usual
temperature.  Hence, the temperature $T$ of the hard sphere gas should be
interpreted differently.  One way to relate this temperature to the vibrational
strength of the granular bed is to compare the kinetic expansion of
the granular bed to the thermal expansion of the hard spheres, as was done in
ref.[10].  If we denote by $\bar h(\Gamma)$
the jump height of a single ball in the vibrating bed of
the vibrational strength $\Gamma = A\omega^2/g$ with $A$ and $\omega$ the
amplitude and the frequency of the vibration, then we may set
$$ <\Delta z(T)> = \alpha N D (T/T_c)^2 = \bar h(\Gamma) \eqno (22)$$
from which we can find the relation between the thermal temperature $T$ of the
hard spheres and the vibrational strength $\Gamma$:
$$ \frac{T}{T_c} = (1-\delta) 
\sqrt{\frac{\bar h}{D}\frac{1}{N}\frac{2}{2+2\delta(1-\delta)}}\eqno (23)$$
or, equivalently:
$$ \frac{T}{mg} = \delta\sqrt{\frac{2\bar h DN}{2+\delta(1-\delta)}}
\eqno (24)$$
We point out that
for granular materials excited by vibration
in a two dimensional container, 
$\Delta = mgD/T_c$ was determined by fitting the density profile of ref.[9]
by the Enskog profile.  The estimated value was
$\Delta \approx 4.926$ [3], and the
dimensioness temperature of the vibrating bed was: $T/T_c=0.663$.
However, we point out that we have not
taken into account (a)
the internal degrees of freedom [16] of the macroscopic particles,
such as rotation, and (b) the inelastic collisions,
which may lead to an interesting clustering
instability [17].  Hence, one has to be
somewhat careful in extending the results of elastic hard spheres to
granular materials. 

\vskip 1.0 true cm
\noindent {\bf Acknowledgment}
\vskip 0.5 true cm
The author wishes to thank D. A. Kurtze and H. Hayakawa
for helpful discussions, and Y. Levin
for bringing his attention to refs. [5] and [11] and for comments on the
paper.
\newpage
\noindent {\bf References}
\vskip 0.3 true cm
\noindent [1] D. C. Hong, Physica A {\bf 271}, (1999) 192.

\noindent [2] D. Enskog and K. Sven, Vetenskapsaked Handle. {\bf 63}, 5 (1922);
S. Chapman and T. G. Cowling, The Mathematical Theory of
Nonuniform Gases, Cambridge, London, 1970.

\noindent [3] P.V. Quinn and D. C. Hong, cond-matt/0005196.

\noindent [4] J.G. Kirkwood, J. Chem. Phys. {\bf 7},
(1939) 911.  N. N. Bogolyubov. J. Phys. USSR {\bf 10}, (1946) 257.
M. Born and M. S. Green, ``A General Kinetic Theory of Liquids.''
Cambridge University Press, Cambridge (1949).

\noindent [5] P. Tarazona, Phys. Rev. A {\bf 31}, (1985) 2672; P.
Taranzona, Mol. Phys. {\bf 52}, (1984) 81.

\noindent [6]  J.K. Percus and G.J. Yevick, Phys. Rev. {\bf 110}, (1958) 1.

\noindent [7] Y. Levin, cond-matt/0007288.

\noindent [8] H. Hayakawa and D. C.
Hong, Int. J. Bifurcations and Chaos, Vol. 7, No.5, (1997) 1159.

\noindent [9] E. Clement and J. Rajchenbach, Euro. Phys. Lett. 16, (1991) 133.

\noindent [10] H. Hayakawa and D. C. Hong, Phys. Rev. Lett. {\bf 78}, (1997)
2764.

\noindent [11]  H. Ted Davis, 
{``\it Statistical Mechanics of Phases, Interfaces,
and Thin Films,''} (VHC Publishers, Inc., New York 1996) chapter 4.

\noindent [12] T. Biben, R. Ohbesorge, and H. Lowen, Europhys. Lett. {\bf 71},
(1993) 665.

\noindent [13] D. C. Hong and K. McGouldrick, Physica A {\bf 255}, (1998) 415.

\noindent [14] M. E. Fisher, J. Chem. Phys. {\bf 42} (1965) 3852.

\noindent [15] H. Jaeger, S.R. Nagle, and R. P. Behringer. Rev. Mod. Phys. 
{\bf 68}, (1999) 1259; See also 'Granular Gases', Edited by S. Luding, 
T. Poschel, and H. Herrmann, Springer-Verlag (2000).

\noindent [16] G. Lukaszewicz, Micropolar Fluids: Theory and
Applications (Birkh\``auser, Boston, 1999);
H. Hayakawa, Phys. Rev. E {\bf 61}, (2000) 5477.

\noindent [17] I. Goldhirsch and G. Zanetti, Phys. Rev. Lett. {\bf 70} 
(1993) 1619.

\noindent [18] P. V. Quinn and D. C. Hong, Physica A,274, 572 (1999).
\end{document}